\documentclass[12pt,twoside]{article}
\usepackage{latexsym, hyperref}
\usepackage{subcaption}
\usepackage{graphicx}
\usepackage{caption}
\usepackage{mathrsfs}
 \usepackage{float}
 \usepackage{xcolor} % Allow colors to be defined
 \usepackage{amsmath} % Equations
 \usepackage{amssymb} % Equations
 \usepackage{enumerate} % Needed for markdown enumerations to work
 \usepackage{geometry} % Used to adjust the document margins
 \usepackage{textcomp} % defines textquotesingle
\usepackage{placeins} % stop figures floating.
\usepackage{mathrsfs}
\usepackage{mdframed}

\usepackage{tikz}   %logo

\mdfdefinestyle{RedBox}{%
    linecolor=red,
    linewidth=2pt,
    roundcorner=10mm,
    innertopmargin=5mm,
    innerbottommargin=5mm,
    innerrightmargin=10mm,
    innerleftmargin=10mm,
    backgroundcolor=red!10!white,
    frametitlerule=true,
    frametitlebackgroundcolor=red!20!white }

\mdfsetup{nobreak=true} %suppress page breaks.

\hypersetup{
		breaklinks=true,  % so long urls are correctly broken across lines
		colorlinks = true,
		linkcolor={red}, 
		urlcolor={blue}, 
		citecolor={blue}
		}

\begin{document}
    
\title{\vskip 2mm\bf\Large\hrule height5pt \vskip 4mm
Machine Learning for Quantum Computing Specialists
 \vskip 4mm \hrule height2pt}
\author{Daniel Goldsmith and M M Hassan Mahmud\\[3mm]
Digital Catapult
}
\date{April 2024}
\maketitle
\begin{tikzpicture}[remember picture,overlay]
   \node[anchor=north east,inner sep=0pt] at (current page.north east)
              {\includegraphics[scale=0.1]{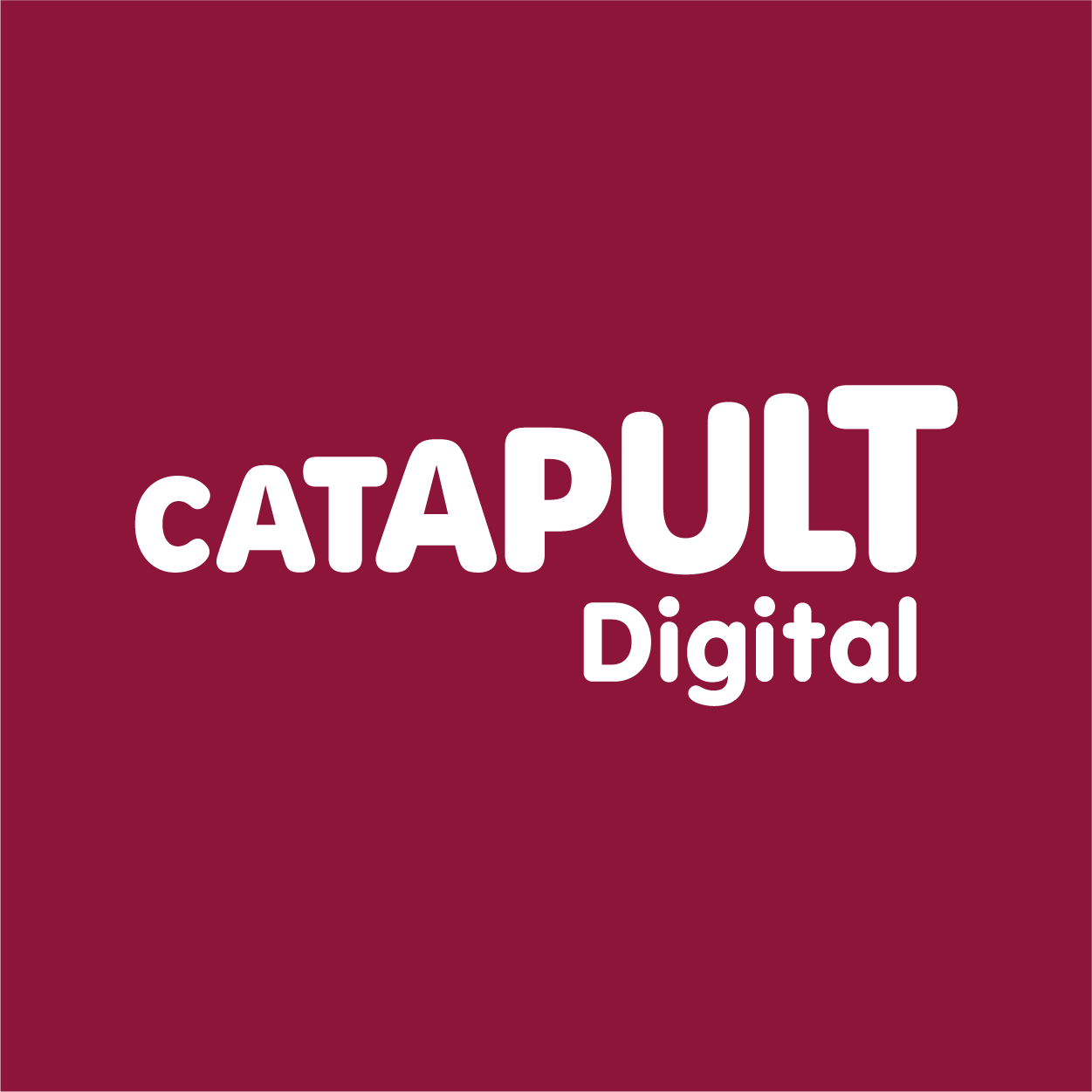}};
\end{tikzpicture}

\begin{abstract}
Quantum machine learning (QML) is a promising early use case for quantum computing.  
There has been progress in the last five years from theoretical studies and numerical 
simulations to proof of concepts.   Use cases demonstrated on contemporary quantum devices include 
 classifying medical images \cite{Mathur2021} and items from the Iris dataset 
\cite{Abbas2020}, classifying \cite{Farhi2018} 
and generating \cite{Huang2021} handwritten images, toxicity screening \cite{Albrecht2022}, 
and learning a probability distribution
 \cite{Benedetti2021}.  Potential benefits of QML include faster training \cite{Abbas2020} 
and identification of feature maps not found classically \cite{Grossi2022}. 
Although,  these examples lack the scale for commercial exploitation, and it 
may be several years before QML algorithms 
 replace the classical solutions, QML is an exciting area.

This article is written for those who already have a sound knowledge of quantum computing and now wish to 
gain a basic overview of the terminology and some applications of classical machine learning 
ready to study quantum machine learning.  
The reader will already understand the 
relevant relevant linear algebra, including Hilbert spaces, a vector space with an inner product.
\end{abstract}

\section{Introduction}

    This review is written for those who already some quantum computing
knowledge and need to gain a basic overview of machine learning before
learning about quantum machine learning (QML).  There has been progress 
in the last five years from theoretical studies and numerical 
simulations to proof of concepts.   Use cases demonstrated on contemporary quantum devices include 
 classifying medical images \cite{Mathur2021} and items from the Iris dataset 
\cite{Abbas2020}, classifying \cite{Farhi2018} 
and generating \cite{Huang2021} handwritten images, toxicity screening \cite{Albrecht2022}, 
and learning a probability distribution
 \cite{Benedetti2021}.  Potential benefits of QML include faster training \cite{Abbas2020} 
and identification of feature maps not found classically \cite{Grossi2022}. 

You may be encouraged to
read that ``a quantum feature map may enable a quantum kernel classifier
on NISQ machines'', but left wondering what a ``feature map'', a
``kernel'' and a ``classifier'' are. You may find it promising that ``a
possible early use case for quantum machine learning is to use quantum
computers to generate the latent spaces for GANs'' and be puzzled
by the terms ``GAN model'', and ``latent space''. This review explores these and other important machine learning concepts at a high level. Key concepts are
highlighted in \textbf{bold} when introduced. This is not an exhaustive
text book on Machine Learning and you are signposted to additional
sources of information to supplement this review. In the interests of
brevity we pass over some popular machine learning algorithms like
\href{https://en.wikipedia.org/wiki/Random_forest}{Random forests} \footnotemark[1]. 

    This review was written by a quantum computing technologist who six
months ago knew that Machine Learning could separate pictures of cats
from pictures of dogs, but not much more. Since then I have been guided
by my expert mentor, Dr M M Hassan Mahmud who has given me great advice and
tuition, and has carefully reviewed this document. You are in safe
hands!
\footnotetext[1]{In this article we adopt a hybrid approach of citing references because of the intended mode of dissemination as an online article. 
For well known concepts from computing, mathematics or quantum physics, we just use an in text hyperlink to an online resource. 
For books or research papers, we include proper citations.}

\subsection{Quantum Computing has given you a good start to machine learning}

    You may already understand most of the mathematics needed for machine
learning from quantum computing. You probably understand
the linear algebra on which much of machine learning is built, because
the same linear algebra is used in quantum computing, albeit with
complex numbers. Most likely you already understand a Hilbert space as a
vector space equipped with an inner product because a quantum state can
be represented as a vector in Hilbert space. The inner product is a key
concept in machine learning, and you have probably already has seen that the inner
product of two quantum states $\psi$ and $\phi$, $\langle \psi | \phi \rangle$, gives a measure of the overlap, or
simularity between the two states. You will understand that a quantum state
can be represented as a vector, and an operation on a quantum state can
be represented as a matrix that produces a new vector representing a
different quantum state. Matrices and vectors are ubiquitous in machine
learing. The symbols for matrices and vectors are shown in \textbf{bold}
in this review.

    You will understand that a matrix \pmb A is Hermitian if $\pmb A =\pmb A^{\dagger}$ , where $\pmb A^{\dagger}$ represents
taking the transpose of the matrix, and then the complex conjugate of
each matrix element. In machine learning the equivalent to the Hermitian
of a matrix is a symmetric matrix, as complex numbers are not normally
relevant. A matrix is symmetric if $\pmb A = \pmb A^T$ where
$\pmb A^T$ represents the transpose of a matrix. We will see that
covariance matrices are symmetric.

    You will understand that a square matrix
$\pmb A \in \mathscr{R}^{n \times n}$ has eigenvalues 
$\lambda_i \in \mathscr{R}$ and corresponding eigenvectors
$\pmb x_i \in \mathscr{R}^{n}$ such that:
\begin{equation}
\pmb A \pmb x_i = \lambda_i \pmb x_i 
\end{equation}
    You will also have some knowledge of probability. If you would like to
brush up on the underlying maths, then we recommend the on-line book
\href{https://mml-book.github.io/book/mml-book.pdf}{Mathematics for
Machine Learning} \cite{MML}. In fact, add a few sections on complex
numbers, and this book could be called \emph{``Mathematics for Machine
Learning and Quantum Computing''}.

    You may also be familiar with \href {https://www.python.org/}{Python}, and may have 
used \href{https://jupyter.org/}{Jupyter} notebooks. Many Machine learning libraries such as
\href{https://pytorch.org/}{Pytorch} and
\href{https://www.tensorflow.org/}{Tensorflow} use Python.  If you have not worked with Python a good introductory book is ``\emph{Python Crash Course"} \cite {Mattes}.  
If you install \href{https://www.anaconda.com/}{Anaconda} then Python and Jupyter will be downloaded as part of the installation.  Jupyter notebooks are straightforward to pick up.  
Python code can be written directly into code cells, and text, including mathematical formulae, 
can be written in \href{https://www.markdownguide.org/cheat-sheet/}{Markdown} cells.  
The code can be changed directly on your computer to gain insight into the example. 

    This review provides links
to Jupyter notebooks hosted on \href{https://github.com/digicatapult/QTAP}
{the GitHub repository for our quantum technology access programme (QTAP)}
 \cite {QTAP} showcasing the comprehensive and easy to learn
\href{https://scikit-learn.org/stable/}{Scikit Learn} machine learning
package. Machine Learning is a practical subject and it is really useful
to work on real examples. \href{https://www.kaggle.com/}{Kaggle} has
over 50,000 public datasets available for free that you can practice on.

\begin{mdframed}[style=RedBox, frametitle={%
\hypertarget{caveats}{%
\subsection{Quantum machine learning - some caveats}
\label{Quantum machine learning - some caveats'}}}]

Most likely you are reading this review because you are interested in
Quantum Machine Learning (QML). We will point out use cases for QML. Be
aware though, that is not a complete review of the current status of
QML. The examples illustrate the text and point to additional references.  Hopefully this will whet your appetite for more.

When reading the literature on QML you need to be aware of the following caveats:

    1.  We are in the age of \href{https://medium.com/@DigiCatapult/is-useful-quantum-computing-possible-in-the-nisq-era-18f423be051a}{Noisy Intermediate Scale Quantum devices}
  (NISQ) with high error rates and few qubits.

    2. In theory these errors can be corrected using \textbf{logical} qubits where a
  quantum state is held across many \textbf{physical qubits} and a
  corrupted physical qubit can be detected and corrected by referring to
  the other uncorrupted physical qubits, but there are significant challenges to using quantum error correction for useful computations.

     3.  At some stage in the future the expectation is that we will have the \textbf{fault tolerant}
  quantum computers with large numbers of error corrected logical qubits predicted by Preskill \cite {Preskill1997}. Increasing the number of physical qubits and getting error correction working
  properly will take at least a decade, and probably much longer.

    4.  There is no commercially available \href{https://en.wikipedia.org/wiki/Quantum_memory}{quantum memory} to store quantum states, 
meaning that it is difficult to carry out quantum calculations that take longer than the short qubit decoherence times. Without quantum memory, converting machine
  learning data into quantum states will be very difficult on today's noisy devices.

  5.  Many quantum computing algorithms, such as Grover's search algorithm \cite{Grover1997},
  the Harrow Hassidim Lloyd (HHL) \cite{Harrow2008} algorithm for matrix inversion and
  Quantum Phase Estimation \cite{Kitaev1995}, that might be useful for QML, are not feasible until fault tolerant
  quantum computers become widely available.

\end{mdframed}

\section{What is Artificial
Intelligence?}
    \textbf{Artificial Intelligence} (AI) dates back to the earliest days of
computing. AI can be defined as ``\emph{the set of technologies that
solve problems that require human like intelligence}''. What is regarded
as ``\emph{human like intelligence}'' ratchets up over time, and as we
will see, much deployed AI is now just regarded as software. So the
unofficial definition of AI is ``\emph{anything that humans can do but
computers cannot do \ldots{} yet!}''.

    The book \href{https://zoo.cs.yale.edu/classes/cs470/materials/aima2010.pdf}{\emph{Artificial
Intelligence, A Modern Approach}} \cite {AI} gives a good overview of AI.
Our colleague at Digital Catapult, Dr Robert Elliot Smith, has written
an important book \href{https://www.rageinsidethemachine.com/}{``\emph{Rage Inside The Machine}''} \cite{Rage}. As well as
highlighting the involvement of AI algorithms in driving intolerance,
Rob Smith puts AI and ML into their historical context, which I found
valuable. Another very useful book is \href{https://link.springer.com/book/10.1007/978-3-030-83098-4}{``\emph{Machine Learning with
Quantum Computers}''} by Maria Schuld and Francesco Petruccione \cite{Schuld}.
This books covers both machine learning and quantum computing, and looks
at quantum machine learning algorithms likely to be useful in the
future. Maria Schuld has posted insightful videos on YouTube, for
example, \href{https://www.youtube.com/watch?v=GEcIDVQaNp4}{Is quantum
advantage the right goal for QML?}

    I also found the book ``\emph{Machine Learning for Absolute Beginners}''
\cite{Dummy} by Oliver Theobald useful as it gave a good basic overview of ML.

    AI has its roots in computer science and draws from diverse fields including logic, probability theory, statistics, neuroscience, information theory, 
cognitive science, linguistics and even biology.  

    \textbf{Machine Learning} is a subset of AI, and machine learning
algorithms have the ability to learn from patterns in data without being
explicitly programmed. There is a useful MIT OpenCourseWare Introduction
to Machine Learning available on
\href{https://www.youtube.com/watch?v=h0e2HAPTGF4}{YouTube}.

    A further subset of machine learning is \textbf{Deep Learning}, in which
\textbf{artificial neural networks} adapt to learn from vast amounts of
data. Deep learning methods have had tremendous success in the real world in recent years, and a lot of the current excitement around AI/ML can be attributed to their success.  There is another great MIT OpenCourseWare Introduction to Deep
Learning also available on
\href{https://www.youtube.com/watch?v=h0e2HAPTGF4}{YouTube}.

 The relationship between the different components of Artificial Intelligence is shown in Figure \ref{fig:1}:

\begin{figure}[h]
    \centering
    \includegraphics[width=0.8\textwidth]{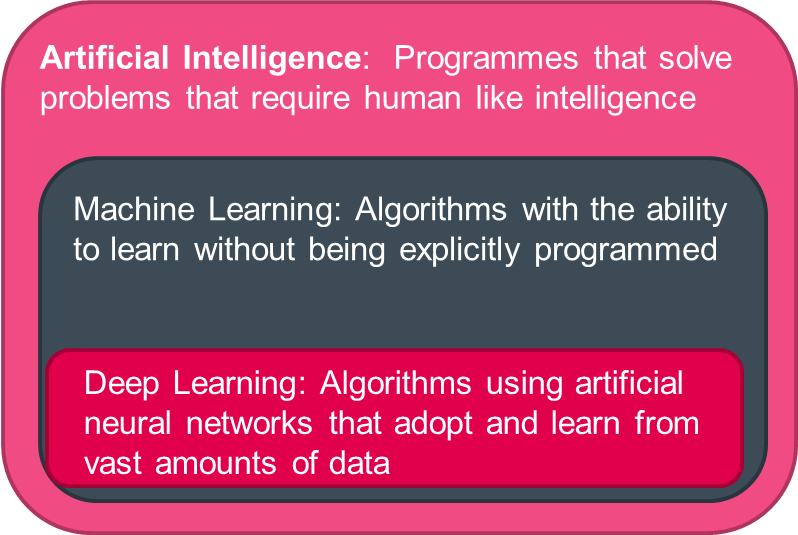}
    \caption{How Artifical Intelligence, Machine Learning and Deep Learning fit together}
    \label{fig:1}
\end{figure}

\hypertarget{expertsystem}{%
\subsection{My first experience of AI - the VAT ``expert
system'}\label{My first experience of AI - the VAT ``expert
system'}}

\begin{figure}[h]
    \centering
    \includegraphics[width=1\textwidth]{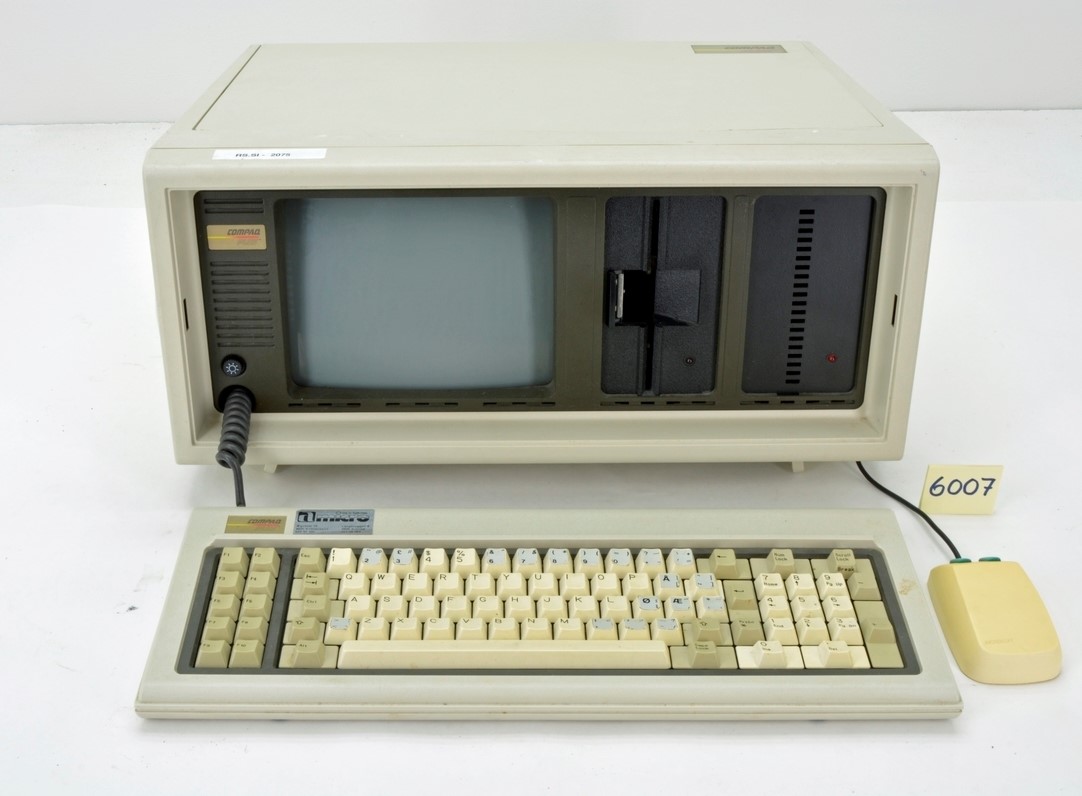}
    \caption{The COMPAQ microcomputer - Thanks to the Norsk Tekisk Museum CC BY-SA 3.0}
    \label{fig:2}
\end{figure}

    When I was working as a auditor in the early nineties we used an AI VAT
\textbf{``expert system''} to give our clients VAT advice. We
interviewed the client, and asked pre-defined questions like ``do you
use cars for business purposes?'', ``what is your turnover?'', input the
answers into our COMPAQ microcomputer shown in Figure \ref{fig:2} and printed out the
results on a dot matrix printer to give to the client.

    With hindsight our ``expert system'' suffered from all the issues of
early expert systems identified by Rob Smith in his book \cite {Rage}. It needed a lot
of time from very expensive VAT and computing experts to design and
build. It was fragile, and quite often the carefully studied output
simply advised the client to consult with human experts.

    This ``expert system'' needed to be explicitly programmed for each and
every situation it would encounter. In the next example we will consider
a simple \textbf{machine learning model} that \textbf{learns} from a
pattern in the data to predict values for unseen data.

    A \textbf{machine learning model} consists of an algorithm which has been optimised on training data to make useful predictions of unseen data.

    \hypertarget{an-introductory-example---linear-regression}{%
\subsection{An introductory example - linear
regression}\label{an-introductory-example---linear-regression}}

    I have placed a
\href{https://github.com/digicatapult/QTAP/blob/main/Python/ML/Linear_Regression.ipynb}{Jupyter
Notebook} on GitHub demonstrating an example of linear regression using
the Scikit Learn package. This is an example of \textbf{supervised
learning}, where the algorithm predicts values based on examples
provided. Although relatively straightforward, this example illustrates
many of the points relevant for more complex machine learning models.
This algorithm pre-dates computers, since for small data sets the
calculations can be done by hand.

    \hypertarget{Aim of the linear regression example}{%
\subsubsection{Aim of the linear regression example}\label{Aim of the linear regression example}}

    The aim of this basic machine learning example is to predict the company turnover from the number of employees.

    \hypertarget{loading-the-data-using-pandas}{%
\subsubsection{Loading the data using
pandas}\label{loading-the-data-using-pandas}}

    In my Notebook \href{https://pandas.pydata.org/}{Pandas}, a Python data
analysis tool, is used to load and process the dataset: a list of UK
quantum technology companies with the company turnover, numbers of employees and
latitudes and longitudes. The dataset is loosely based on real UK data.
A plot of employees against turnover from the dataset is shown in Figure \ref{fig:3}:

\begin{figure}[h]
    \centering
    \includegraphics[width=1\textwidth]{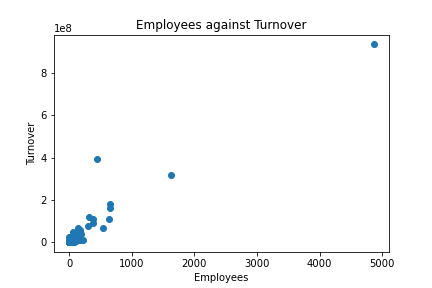}
    \caption{Plot of employee data against company turnover: the raw data}
    \label{fig:3}
\end{figure}

    \hypertarget{machine-learning-terminology-used-to-describe-the-data}{%
\subsubsection{Machine Learning terminology used to describe the
data}\label{machine-learning-terminology-used-to-describe-the-data}}

    The company turnover, numbers of employees, latitude and longitude are
called \textbf{features}. Each line of the file is called a
\textbf{sample}. Each sample can be regarded as a 4 dimensional vector,
with one dimension for each feature, allowing concepts from linear
algebra to be used. The set of all feature vectors is know as the
\textbf{design matrix}. In this example the latitudes and longitudes are
not used. The company turnover is a \textbf{label} in this example and the number of employees is
an \textbf{input}. In any data analysis ``Garbage In, Garbage Out''
applies, so the data is first listed, plotted on a graph to check for
outliers and any records with unavailable data are removed before
proceeding. Whilst data analysis, and \href{https://en.wikipedia.org/wiki/Data_science}{data science} more generally, are very important to machine learning in practice,
it is too large a subject to be considerd further in this review.

    \hypertarget{training-and-test-data}{%
\subsubsection{Training and test data}\label{training-and-test-data}}

    Next, the data is split into two components, \textbf{training data}
which makes up 70\% of the data, and \textbf{test data}, which makes up
the remaining 30\%. The \textbf{model} is \textbf{trained} with the
training data, and then makes predictions which are validated against the
\textbf{test data}, showing how good the model is at
\textbf{generalising} from the training data to the test data, a key
feature of all models.

    More complicated models need \href{https://en.wikipedia.org/wiki/Training,_validation,_and_test_data_sets}{validation data, as well as training and test data}.

    \hypertarget{the-hypothesis-space}{%
\subsubsection{The hypothesis space}\label{the-hypothesis-space}}

    Every machine learning problem has a family of candidate solutions chosen by the ML engineer.  Each family of solutions, for example,  
polynomial functions, trigonometric functions and linear functions, is called a \textbf{hypothesis space}.  
The set of hypothesis spaces are limited only by our imagination.  
The aim of a particular ML algorithm is to use the training data to find the best solution in a given hypothesis space.

There is no way of being sure which is the
best hypothesis space and algorithm to use. In fact, the \textbf{``no free lunch''} theorem
implies that there is no single best machine learning algorithm. All
algorithms work well in some cases and badly in other.

    In this example, if we write \(x\) as the number of employees and
\(\hat y\) as the predicted turnover, then the hypothesis spaces are different families of \textbf{predictor functions} that returns a prediction $\hat y = f(x)$ from an input \(x\).  

    \hypertarget{the-inductive-bias}{%
\subsubsection{The inductive bias}\label{the-inductive-bias}}

\begin{figure}[h]
    \centering
    \includegraphics[width=1\textwidth]{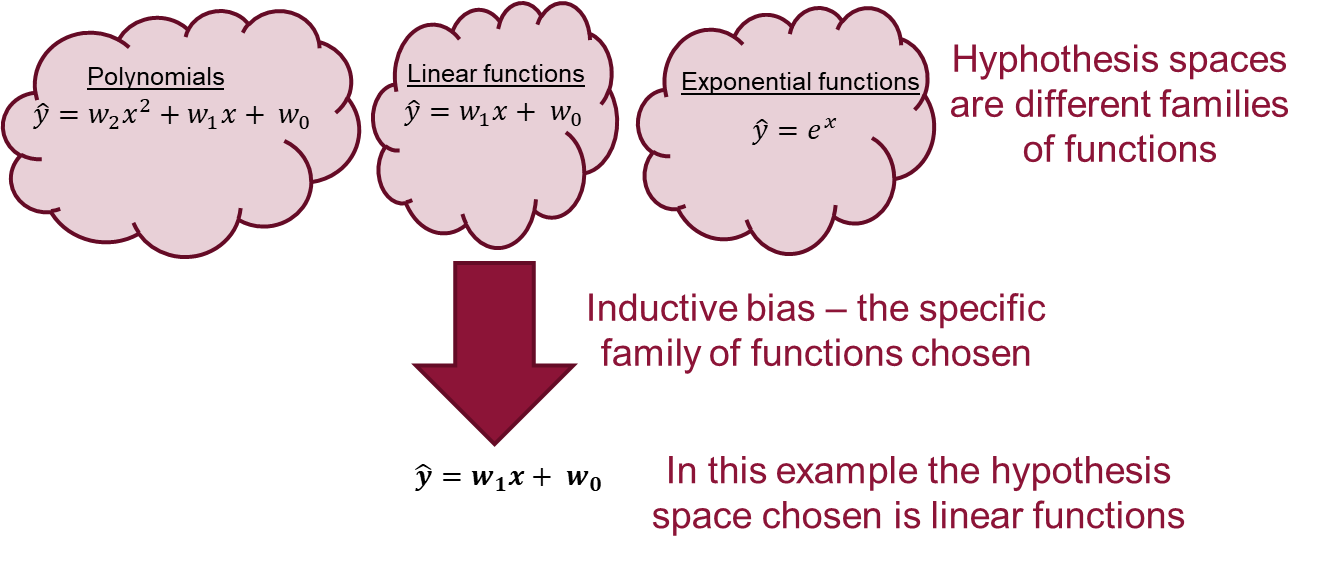}
    \caption{The inductive bias identifies the hypothesis space to be used from all possible hypothesis spaces}
    \label{fig:4}
\end{figure}

To get a tractable ML algorithm it is necessary to
make specific assumptions about the form of the mapping from inputs to labels as shown in Figure \ref{fig:4}. These
assumptions are called the \textbf{inductive bias}. The hypothesis space
encodes these assumptions mathematically. Just as we all have
\textbf{sub-conscious biases} which means that we may make incorrect
decisions without realising it, we may also incorrectly tie our models
to a mistaken assumption. 

    In this example it is assumed that the
predictor function is linear, with \(f(x) = w_1 x + w_0\). The model
needs to estimate \(w_1\), the slope of the line, and \(w_0\) the
offset.

    It is normal in ML to regard the parameters of a model as a vector
\(\pmb \theta\). Here \(\pmb \theta = [w_1, w_0]^T\). I have followed
the common convention in ML of writing a column vector as the transpose
of the row entries, which saves space. The predictor function returns
the prediction from an input \(\pmb x\) and parameters \(\pmb\theta\)
and is writen as \(f(\pmb x,\pmb \theta)\).

\hypertarget{choosing-and-optimising-a-loss-function}{%
\subsubsection{Choosing and optimising a loss
function}\label{choosing-and-optimising-a-loss-function}}

\begin{figure}[h]
    \centering
    \includegraphics[width=1\textwidth]{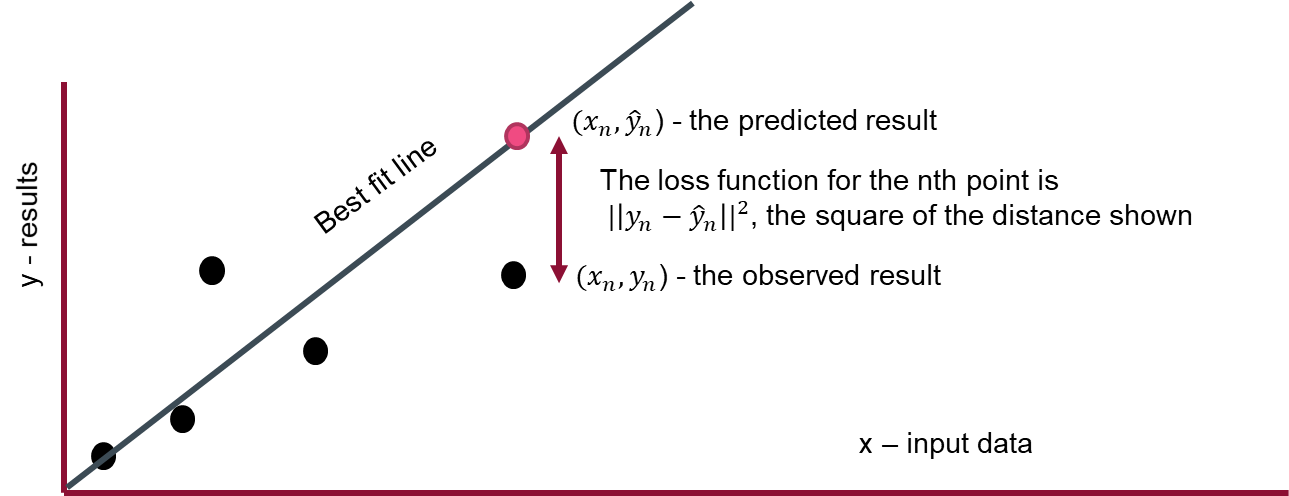}
    \caption{The loss function shown diagramatically}
    \label{fig:5}
\end{figure}

    Once we have chosen our hypothesis space we need to find the best parameters, or hypothesis, to use.  To find the best parameters we need a way to decide whether one hypothesis is better than another.  This is done through a \textbf{loss function} which measures the quality of prediction made by our ML model.  The loss function \(\mathscr{L}(\pmb\theta\)) determines the usefulness of the predictions, for example, in estimating turnover.  So, choosing a loss function is a crucial decision when setting up a ML problem.  In this example, the loss function for each sample is chosen as the square of the distance between the actual turnover \(y\) and the predicted turnover 
\(\hat y\): \[\mathscr{L}(y, \pmb \theta) = ||y - \hat y||^2 = ||y - f(x,\pmb\theta)||^2\]The
loss function is shown in Figure \ref{fig:5}

    The parameters are chosen to minimise the sum of the loss function over
all the samples in the training set, $$\mathscr{L}(\pmb\theta) = \sum_i \mathscr{L}(y_i, \pmb\theta)$$ which is a ``least squares'' \textbf{optimisation} problem. There is an exact solution under certain conditions.

\begin{figure}[h]
    \centering
    \includegraphics[width=1\textwidth]{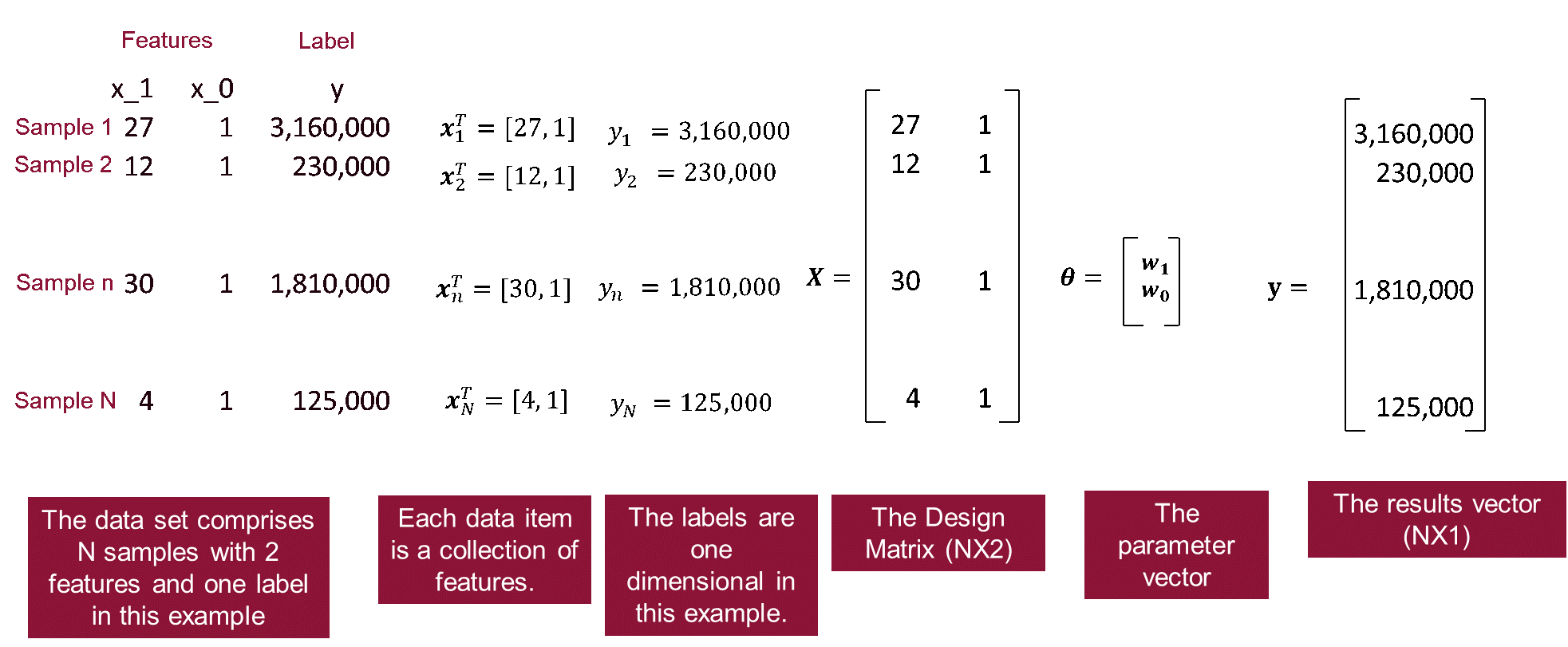}
    \caption{This diagram show the dataset with features and labels, the
results vector, design matrix, and the parameter vectors}
    \label{fig:6}
\end{figure}

    If we consider the nth sample we can write \(x_n\) as a vector
\(\pmb x_n\). Here \(\pmb x_n = [x_n, 1]^T\).  This means that we can neatly write \(f(x_n) = \pmb x_n^T \pmb \theta\), the vector dot product, or inner product, of the input vector and the
parameter vector. If there are N samples we can write the \textbf{design matrix} as
\(\pmb{X} = [\pmb x_1, .... ,\pmb x_n, ...\pmb x_N]^T \in \mathscr{R}^{N \times 2}\),
and the observed results vector
\(\pmb{y} = [y_1,.... ,y_i,....y_N]^T \in \mathscr{R}^{N}\). All these
different matrices and vectors are illustrated in Figure \ref{fig:6}

Using this notation it is shown in Chapter Nine of
\href{https://mml-book.github.io/book/mml-book.pdf}{Mathematics for
Machine Learning} \cite {MML} that by differentating \(\pmb\theta\) a minimum of the
loss function is found when
\[{\pmb\theta} = (\textbf{X}^T\textbf{X})^{-1}\textbf{X}^T\textbf{y}\].

    In other words this is a matrix inversion problem where an equation of
form \(\pmb{A}\pmb{\theta} = \pmb{b}\) with
\(\pmb{A} = \pmb{X}^T\pmb{X}\) and \(\pmb{b} = \pmb{X}^T\pmb{y}\) needs
to be solved. Figure \ref{fig:7} below shows the dataset, and the best fit line for the training data against the test
data calculated by SciKit Learn:

\begin{figure}[h]
    \centering
    \includegraphics[width=0.8\textwidth]{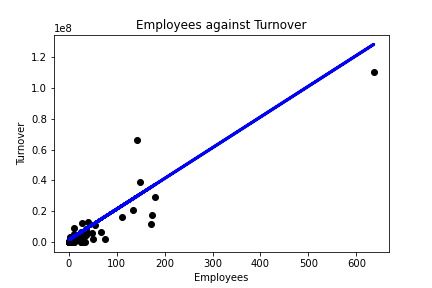}
    \caption{Plot of employee data against company turnover: a best fit
straight line from the training data against the test data}
    \label{fig:7}
\end{figure}

    \hypertarget{useful-linear-regression-documentation}{%
\subsubsection{Useful linear regression
documentation}\label{useful-linear-regression-documentation}}

    Details of the LinearRegression class used are found on the SciKit Learn
(abbreviated to sklearn)
\href{https://scikit-learn.org/stable/modules/generated/sklearn.linear_model.LinearRegression.html}{documentation}.
There are notes on linear models in the SkiKit Learn
\href{https://scikit-learn.org/stable/modules/linear_model.html\#ordinary-least-squares}{user
notes}.
\begin{mdframed}[style=RedBox, frametitle={%
    \hypertarget{qml-application}{%
\subsubsection{QML application - linear regression}\label{qml-application}}}]
    The matrix inversion problem above could, in principle, be solved by the quantum HHL algorithm. However, this is likely to require fault tolerant quantum
computers, so the \hyperlink{caveats}{caveats} above are relevant.
\end{mdframed}

    \hypertarget{overfitting}{%
\subsection{Overfitting}\label{overfitting}}

\begin{figure}[h]
    \centering
    \includegraphics[width=1\textwidth]{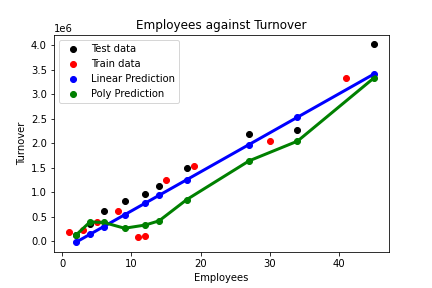}
    \caption{Plot of employee data against company turnover : the polynomial has
over-fitted the data}
    \label{fig:8}
\end{figure}

    In the example above we predicted unseen data points using a best fit
straight line \(f(x) = w_1 x + w_0\). We might think that we could
increase the accuracy of our prediction by changing our
\textbf{inductive bias} to use a different function from a different \textbf{hypothesis space}, a polynomial:
\[ f(x) = w_6 x^6 + w_5 x^5 + w_4 x^4 + w_3 x^3 + w_2 x^2 + w_1 x + w_0\]This
predictor function has more parameters, and so might fit the dataset  better.
In the second
\href{https://github.com/digicatapult/QTAP/blob/main/Python/ML/Overfitting.ipynb}{example}
I have done exactly that, using some manufactured data. However, there
is a snag. Now, the best-fit curve shown in Figure \ref{fig:8} has too many parameters: it fits
the training data well, but does not generalise to the unseen test data
as well as the simple straight line.  This is called \textbf{over-fitting} For some data-sets the fit to the polynomial might give better
generalisation than the straight line, but here the performance is
worse.

    One  solution to overfitting is \textbf{regularisation}. Early stopping, feature selection 
and ensembling are other solutions widely used.  With regularisation the
predictor function could be simplified, using a straight
line, not a polynomial. Alternatively a penalty term
\(\lambda||\pmb\theta||^2\) can be applied to the loss function. This
works because the magnitude of the parameter values can become larger
with overfitting.

    \hypertarget{clustering}{%
\subsection{Clustering}\label{clustering}}

    The examples so far have been examples of \textbf{supervised
learning}: we have fed our model training examples, with input values
and labels (employee numbers), and then our model has been able to
predict the label for unseen data.

    If an AI system is to learn about its environment, it must derive
patterns in the data without any examples. This is called
\textbf{unsupervised} learning, and is often very useful for exploratory data analysis. An example of this is
\textbf{clustering}.

    In the third
\href{hhttps://github.com/digicatapult/QTAP/blob/main/Python/ML/Clustering.ipynb}{example}
we use \textbf{clustering} to investigate the latitudes and longitudes
of the UK quantum technology companies that we met first in the linear
regression example.

    \hypertarget{Aim of the clustering example}{%
\subsubsection{Aim of the clustering example}\label{Aim of the clustering example}}

The aim of the clustering example is to group the companies into clusters so that companies in the same cluster are close together.

    \hypertarget{K-Means clustering}{%
\subsubsection{K-Means clustering}\label{K-Means clustering}}

\begin{figure}[h]
    \centering
    \includegraphics[width=0.8\textwidth]{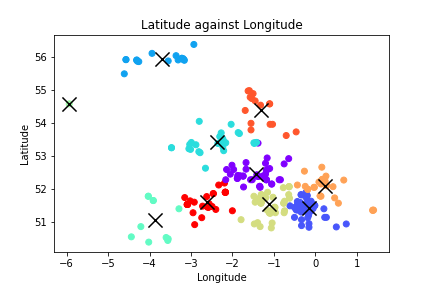}
    \caption{Plot of UK quantum technology company locations, and their K-means
centroids, with K = 10}
    \label{fig:9}
\end{figure}

    A tool called
\href{https://scikit-learn.org/stable/modules/generated/sklearn.cluster.KMeans.html}{K-Means
clustering} is used to group the data into 10 clusters, as shown in Figure \ref{fig:9}. In this
algorithm \(K = 10\) initial centroids are chosen randomly. Data points
are assigned to clusters so that the \textbf{distance} from data points to the
initial centroids is minimised. Then \(K = 10\) new centroids are chosen
at the centre of the new clusters. The data points are reassigned, and
the process continues iteratively, until there are no further
re-assignments of data points to new clusters.

    A key point is that the \textbf{``distance''} is the \textbf{Euclidean
distance} in two dimension, where the distance between vectors
\(\pmb a = (a_x, a_y)\) and \(\pmb b = (b_x, b_y)\) is
\(\sqrt{(a_x - b_x)^2 + ((a_y - b_y)^2}\).

    The \textbf{``centroids''} of the clusters found are shown as crosses.

\begin{figure}[h]
    \centering
    \includegraphics[width=0.8\textwidth]{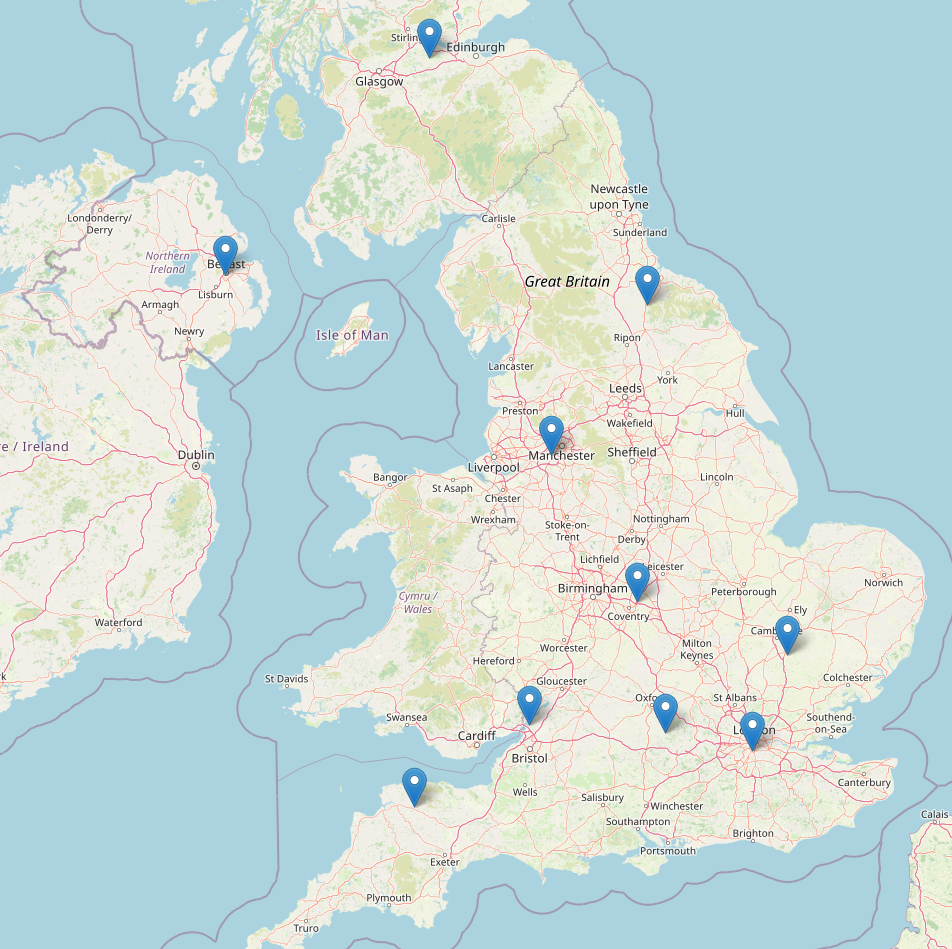}
    \caption{Plot of the K=10 Means centroids for UK quantum technology
companies on a map of the UK}
    \label{fig:10}
\end{figure}

    If we look at the plot of these centroids on a map of the UK shown in Figure \ref{fig:10} we see that
the quantum technology companies are clustered around Quantum Hubs at
Birmingham, Glasgow, Oxford and important Universities at Cambridge,
London, Bristol, Belfast, Liverpool and Manchester. There are also
companies in the South West and North East. The machine learning
algorithm has found these clusters without any external help.

    There is more information on
\href{https://scikit-learn.org/stable/modules/clustering.html}{SciKit
learn}. If you would like to practice your clustering there are some
artificial and real world datasets provided by
\href{https://github.com/deric}{Thomas Barton} available on
\href{https://github.com/deric/clustering-benchmark}{GitHub}.

\FloatBarrier

    \hypertarget{increasing-dimensionality---the--trick}{%
\subsection{Increasing dimensionality - the kernel
trick}\label{increasing-dimensionality---the-kernel-trick}}

The next example is a \textbf{binary classification} supervised learning
problem. I have placed a
\href{https://github.com/digicatapult/QTAP/blob/main/Python/ML/Linear Support Vector.ipynb}{Jupyter
Notebook example} on GitHub. 

    \hypertarget{Aim of the support vector classifier example}{%
\subsubsection{Aim of the support vector classifier example}\label{Aim of the support vector classifier example}}

The aim of a support vector classifier is to predict labels for unseen data.  The input is a two dimensional co-ordinate vector, the labels are binary \(y_i \in\{-1 ,1\}\), and as explained below, the data points can be separated by a linear threshold. 

    \hypertarget{SVC}{%
\subsubsection{The support vector classifier}\label{The support vector classifier}}
 
\begin{figure}[h]
    \centering
    \includegraphics[width=0.8\textwidth]{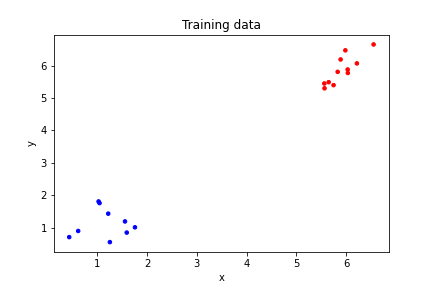}
    \caption{Plot of the training data for the support vector classifier
example}
    \label{fig:11}
\end{figure}

  In Figure \ref{fig:11} different colours have been given to training data points with different labels.

    The algorithm used, the \textbf{support vector classifier}, is based on
an inner product, which gives a measure of similarity between two
vectors. In this example the inner product is defined as the dot product
and the inner product of two vectors \(\pmb x\) and \(\pmb x'\) is:
\[ \langle \pmb x , \pmb x' \rangle = \pmb x^T \pmb x' \]

    An inner product is a map from two vectors in the data domain
\(\mathscr{X}\) to a real number which meets certain
\href{https://en.wikipedia.org/wiki/Inner_product_space}{axioms}.
\[\kappa : \mathscr{X} \times \mathscr{X} \rightarrow \mathscr{R}\]

    Just like the inner product in quantum mechanics, the inner product in
Machine Learning gives a measure of similarity between two vectors.
Using the inner product allows access to geometrical concepts such as
the angle between two vectors, and their length.

\begin{figure}[h]
    \centering
    \includegraphics[width=0.8\textwidth]{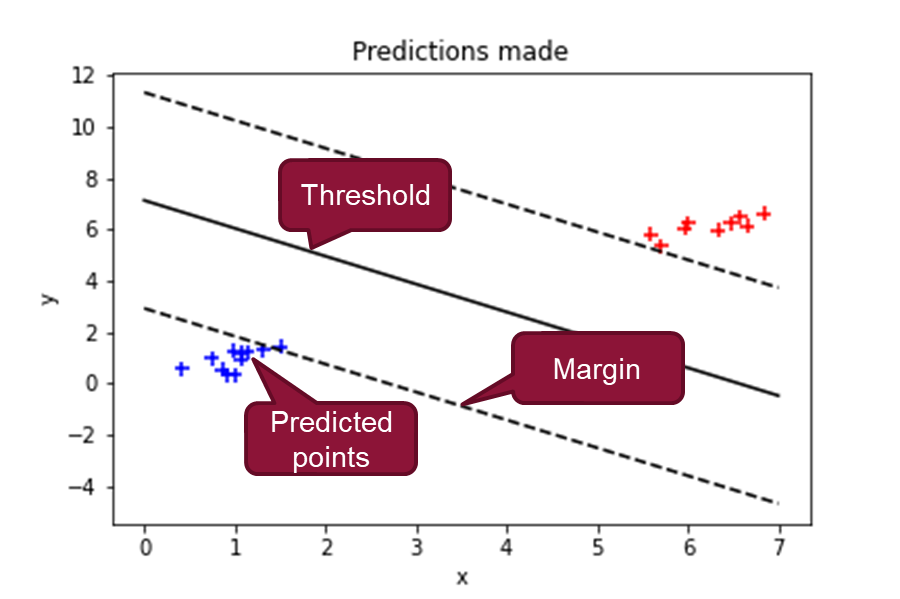}
    \caption{Plot of predictions showing the threshold, and margins}
    \label{fig:12}
\end{figure}

    In this example a \textbf{support vector classifier} uses the dot product to find a \textbf{threshold}, a line to classify the new input data. 
Any data above the threshold line is classified as red, and anything below the
line is classified blue. Figure \ref{fig:12} shows that the support vector
classifier gives accurate predictions for the test data. The graph also
shows the threshold and the \textbf{margin}, the distance between the
threshold and the closest observation to the threshold. The margin is
shown as two dashed lines.

\FloatBarrier

    In support vector classification, the threshold is found by maximising the margin.  The inner product plays a fundamental role in solving this optimisation problem.

    \hypertarget{The-support-vector-machine}{%
\subsection{The support vector
machine}\label{The-support-vector-machine}}

In all the examples so far there is an explicit assumption that similar
points are ``close'' in two dimensional space, either because they are
separated by a small Euclidean distance, or have a normalised inner
product for the data position vectors close to one.

\begin{figure}[h]
    \centering
    \includegraphics[width=1\textwidth]{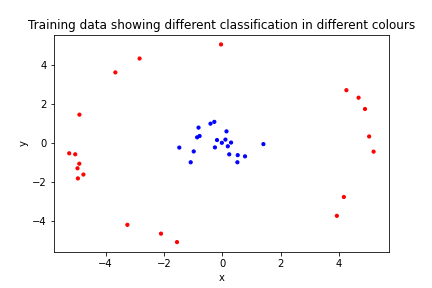}
    \caption{Plot of training data for the support vector machine example}
    \label{fig:13}
\end{figure}

    In the next example similar points are not necessarily close in two dimensions. I
have placed a
\href{https://github.com/goldsmithdaniel/QuantumMachineLearning/blob/main/Kernel\%20example.ipynb}{Jupyter
Notebook} on GitHub illustrating the \textbf{Kernel trick}, which is  a general method to solve classification problems where the data points can't be separated by a linear classifier, as shown in Figure \ref{fig:13}.

    \hypertarget{Aim of the support vector machine example}{%
\subsubsection{Aim of the support vector machine example}\label{Aim of the support vector machine example}}

The aim of the support vector machine example is to separate data points that can't be separated by
a linear classifier.

    \hypertarget{The kernel trick}{%
\subsubsection{The kernel trick}\label{The kernel trick}}

    As explained in ``\emph{Machine Learning with Quantum Computers}'' by
Maria Schuld and Francesco Petruccione \cite {Schuld}  the \textbf{support
vector machine} algorithm requires a \textbf{kernel}, an inner product
which gives a measure of similarity between two vectors, but in this
case the inner product is between two vectors in a higher dimension
feature space \(\mathscr{F}\) defined by a feature map
\(\phi : \mathscr{X} \rightarrow \mathscr{F}\) The kernel \(\kappa\) for
two vectors \(\pmb x\) and \(\pmb x'\) is defined as:
\[\kappa (\pmb x,\pmb x') = \langle \phi(\pmb x) , \phi(\pmb x') \rangle_\mathscr{f}\]

    In this example a feature map \(\phi\) is defined as
\[\phi((x_1, x_2)^T = (x_1, x_2, 0.5 * (x_1^2 + x_2^2))^T\]

\begin{figure}[h]
    \centering
    \includegraphics[width=1\textwidth]{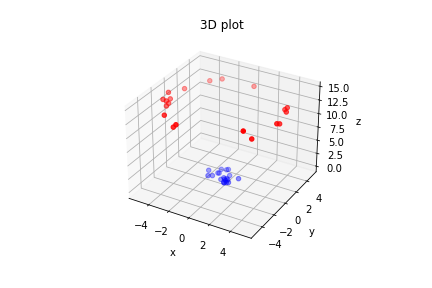}
    \caption{Plot of data in three dimensions}
    \label{fig:14}
\end{figure}

    The feature map is used to transform the dataset from a two dimensional
plane into three dimensional space, as shown in Figure \ref{fig:14}. The points are now
separable. The threshold is a two dimensional plane perpendicular to the
z-axis.

\begin{figure}[h]
    \centering
    \includegraphics[width=1\textwidth]{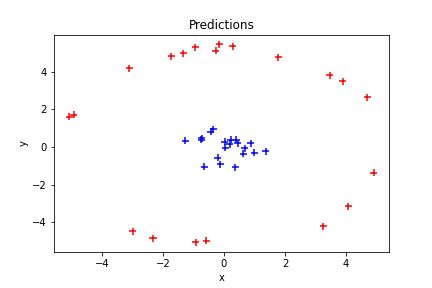}
    \caption{Plot of predictions}
    \label{fig:15}
\end{figure}

 It is not necessary to actually do a mapping and hold the information in
the extra dimensions introduced. All that is needed is to compute the
kernel function between two points \(x\) and \(x'\) and return the
result, which is a real number. This is called the \textbf{kernel
trick}. This can be seen in the Notebook where a new kernel function has
been written to be called by the SciKit Learn support vector classifier,
which does not ``know'' about the new diminension introduced. The
accurate predictions by the support vector machine classifier are shown
in Figure \ref{fig:15}.

    If we added more dimensions to our example the threshold would become a
\textbf{hyperplane} in several dimensions.  There is a great
\href{https://www.youtube.com/watch?v=efR1C6CvhmE}{YouTube video} on
Support Vector machines.

\begin{mdframed}[style=RedBox, frametitle={%
    \hypertarget{the-quantum-kernel-classifier}{%
\subsubsection{The quantum kernel
classifier}\label{the-quantum-kernel-classifier}}}]

    There is a link between the kernel function and the quantum inner
product. The kernel function is defined as:
$$\kappa (\pmb x,\pmb x') = \langle \phi(\pmb x) , \phi(\pmb x') \rangle_\mathscr{f}$$
This is very similar to the quantum inner product:
$$\langle \phi(\pmb x) | \phi(\pmb x') \rangle$$ 
\end{mdframed}

    \hypertarget{the-quantum-kernel-classifier-NISQ}{%
\subsubsection{The quantum kernel
classifier in the NISQ era}\label{the-quantum-kernel-classifier-NISQ}}
    The quantum inner product can
be estimated by a simple circuit, and as mentioned in the introduction,
``a quantum feature map may enable a quantum kernel classifier on NISQ
machines''. However, there are still considerable challenges in
translating the classical data to a quantum form.  Maria Schuld has written an insightful \href{https://arxiv.org/abs/2101.11020}{paper} 
about quantum kernel methods \cite{Schuld2021}. 

Useful progress has been made on NISQ machines.  
A 2022 \href{https://arxiv.org/abs/2211.16337}{paper} by Boris Albrect and others \cite{Albrecht2022} 
describes a quantum kernel running on a 32 qubit neutral atom quantum device.  
In this toxicity screening experiment, molecules were labelled as either
toxic or not toxic, and the QML algorithm classified unseen molecules, with input data being a graph 
representing the arrangement of atoms in the molecule.  They achieved results comparable to the best 
classical algorithms.  

The quantum kernel used is explained in a 2021
\href{https://arxiv.org/pdf/2107.03247.pdf}{paper} by Louis-Paul Henry and others \cite{Henry2021}.
Each atom in the molecules screened was represented by a qubit, and the graph of the molecular structure determined which qubits were allowed to interact.  The quantum state of the system was updated by applying a series of Hamiltonians.  
The precise form of the Hamiltonian pulse sequence was found through training. 
 An observable was measured at the end of the updates, yielding a bit string.  From this bit string a probability distribution $\mathscr{P}(\mathscr{G})$ was built up for each graph by repeating the measurements several times.  

   Once probability distributions had been obtained for each graph, the kernel of two graphs $\mathscr{G}$ and $\mathscr{G'}$ 
was calculated classically as:
$$\kappa(\mathscr{G}, \mathscr{G'}) = exp[JS (\mathscr{P}, \mathscr{P'})] \in {[0.5, 1]}$$ 

  $JS$ is the Jensen-Shannon divergence,
a measure of the distance between the two probability distributions $\mathscr{P}$ and $\mathscr{P'}$.
This kernel was then used in a support vector machine, as described above.    

In a 2018 \href{https://arxiv.org/pdf/1804.11326.pdf}{paper} Vojtech Havlicek and others \cite {Havlicek2018} describe using 5 qubit super-conducting quantum devices to experimentally realise two other quantum kernels.

    \hypertarget{PCA}{%
\subsection{Reducing problem dimensionality - Principal Component
Analysis}\label{reducing-problem-dimensionality---principal-component-analysis}}

\begin{figure}[h]
    \centering
    \includegraphics[width=0.6\textwidth]{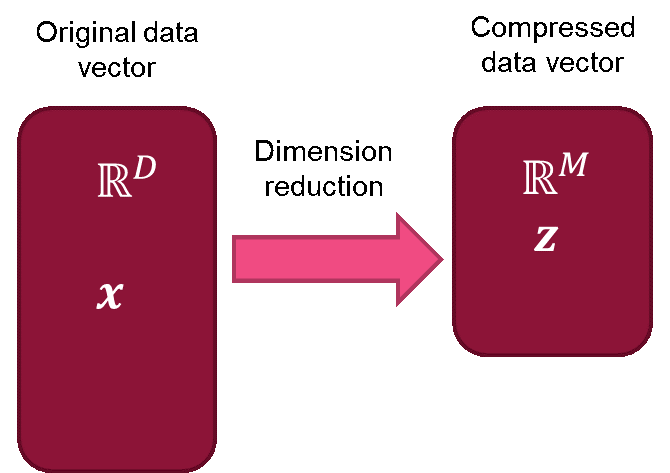}
    \caption{In Principal Component Analysis the original data is compressed,
retaining the key information}
    \label{fig:16}
\end{figure}

    In the example above we \textbf{increased the dimensionality} of the
clustering problem to make points linearly seperable and then used a
support vector machine to find the threshold hyperplane. In the next
example we will use \textbf{Principal Component Analysis (PCA)}, an
unsupervised algorithm, that \textbf{decrease the dimensionality} of the
data to find the key features as shown in Figure \ref{fig:16}.  The mathematics used is interesting, but not essential, and \hyperlink{Appendix1}{an Appendix} explains how the co-variance matrix is diagonalised, and the principal components are then identified as the eigenvectors with the largest eigenvalues.

\begin{figure}[h]
    \centering
    \includegraphics[width=0.8\textwidth]{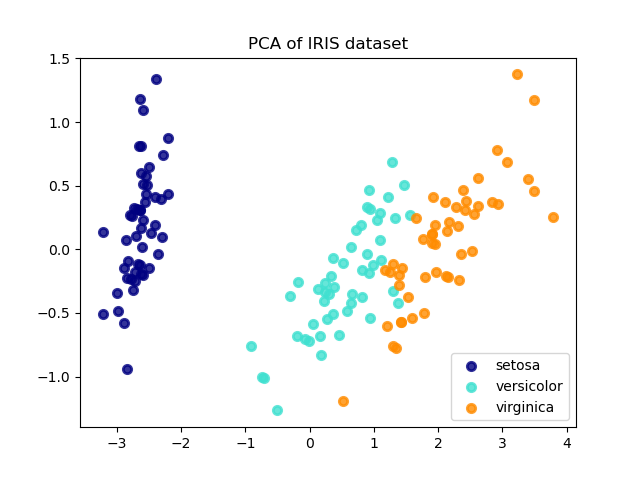}
    \caption{Results of the PCA analysis of the IRIS dataset from SciKit Learn}
    \label{fig:17}
\end{figure}

    If we plot a graph of the data against the principle components then we
can see patterns that we would otherwise miss. An
\href{https://scikit-learn.org/stable/auto_examples/decomposition/plot_pca_vs_lda.html}{example
from SciKit Learn} used the famous IRIS dataset where four features of
the flowers are considered: sepal length, sepal width, petal length and
petal width. When the dimensionality of the problem is reduced to two
dimensions by PCA, it can be seen in Figure \ref{fig:17} that there are three clusters of data,
representing three different species of Iris - Setosa, Versicolour and
Virginica.

    There is more information about PCA on
\href{https://scikit-learn.org/stable/modules/generated/sklearn.decomposition.PCA.html\#sklearn.decomposition.PCA}{SciKit
Learn}.

\FloatBarrier

    \hypertarget{deep-learning}{%
\subsection{Deep learning}\label{deep-learning}}

    Deep learning is inspired by biological models, such as the brain, in
which signals are processed by individual neurons connected to thousands
of other neurons, which ``fire'' dependent on the signals received. In
deep learning these neurons are modelled by networks. A very simple
network is shown in Figure \ref{fig:18}:

\begin{figure}[h]
    \centering
    \includegraphics[width=1\textwidth]{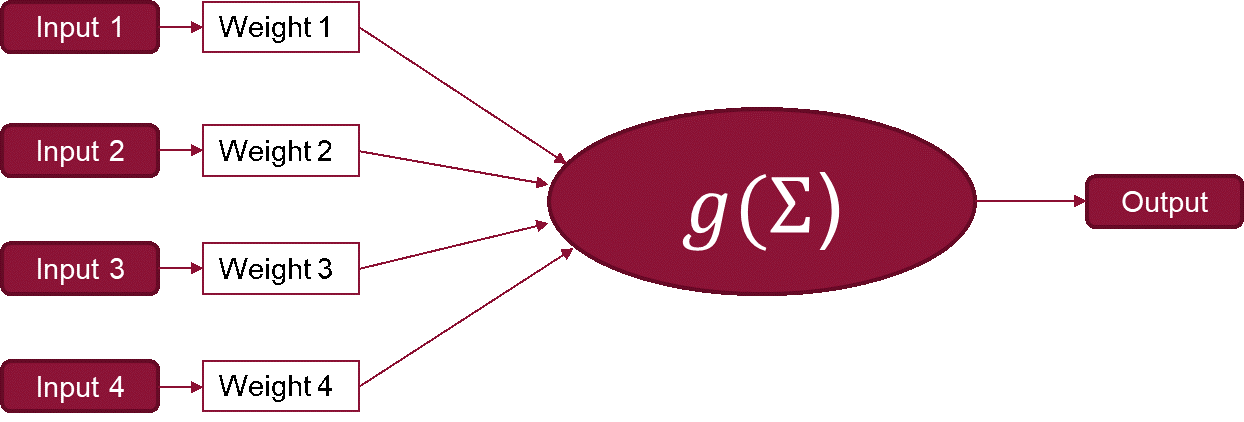}
    \caption{A simple network with four inputs}
    \label{fig:18}
\end{figure}

    In this example there are four inputs and each of these is adjusted by a
weight. The weight adjusted inputs are summed, and a bias
\(b\) is applied. If the input is a vector \(\pmb x\) and the weights
are given by \(\pmb W\) so that the sum \(z\) is given by
$$z = \pmb W^{T} \pmb x + b$$The output will then depend on some
activation function, \(g(z)\). Rectified linear units use the activation
function \(g(z) = max(0,z)\) so there is a linear output if the sum is
greater than zero. Other functions, such as the sigmoid \(S(z)\), which
produces an output between zero and one are also used in some
problems and have an output $$S(z) = \frac {1}{1 + e^{-z}}$$

\begin{figure}[h]
    \centering
    \includegraphics[width=1\textwidth]{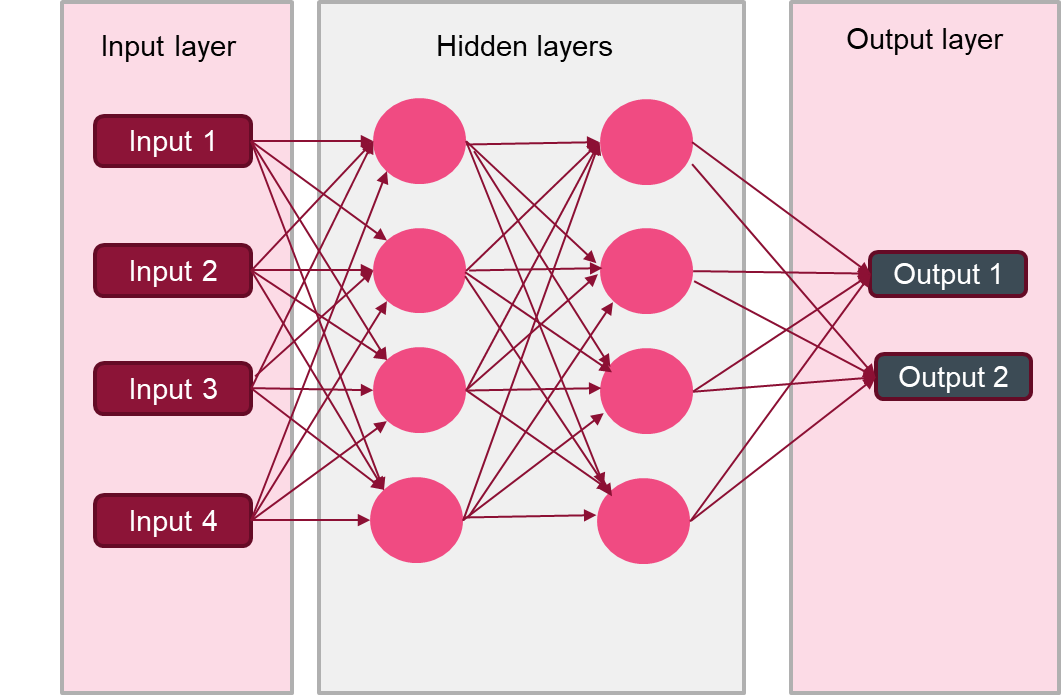}
    \caption{A network with two hidden layers}
    \label{fig:19}
\end{figure}

    The power of deep learning comes from using large numbers of nodes
connected together. In Figure \ref{fig:19} two \textbf{hidden layers} are
added and each arrow is associated with a weight. The idea behind this
is that the hidden layers might enable aspects of the image, like lines,
to be trigger activiations in the first layer, and these are combined so
that geometrical shapes, like triangles, triggger activations in the
second layer, and so on with more abstract levels of detail, until
objects like hand written numbers can be classified.

    In a supervised learning classification problem a loss function is set
to zero if the classification is successful, and to one if it fails.
This loss function is a complex function of the various weights and
activation functions in the model. \textbf{Backpropagation} can be used
to adjust the model weights according to the gradient of the loss
function with respect to the given parameter. A feedback loop can be
used to modify parameters based on the results on previous calculations,
to minimise the loss function.

    Larger versions of these types of networks have been very successful at a wide range of problems including 
complex image and video analysis and generation, solving complex decision making problems at or beyond 
human level capability, natural language processing and generation.

    If you would like to work through an example there is a great
\href{https://pytorch.org/tutorials/beginner/basics/intro.html}{PyTorch
Tutorial} that uses the FashionMNIST dataset to
train a neural network. The model predicts if an input image belongs to
one of 10 classes such as T-shirt, Trouser and Pullover. The tutorial
starts with an introduction to Tensors, which are used to hold a
representation of the network. Then the tutorial walks through loading
and transforming the data, building a neural network, and
\textbf{automatic differentiation} to optimise the model parameters. The
dataset comprises $28 \times 28$ grayscale image and an associated label from one
of the 10 classes. The model has an input has 784 features, one for each
grayscale value, and hidden layers with 512 nodes.

    \hypertarget{GAN}{%
\subsection{Generative Adversarial
Networks}\label{generative-adversarial-networks}}

    You may have read that \textbf{deepfakes} can be generated by Generative
Adversarial Networks, or GANS. A deepfake is a digital object such as an
image, or video clip, produced by Machine Learing techniques. 

\begin{figure}[h]
    \centering
    \includegraphics[width=1\textwidth]{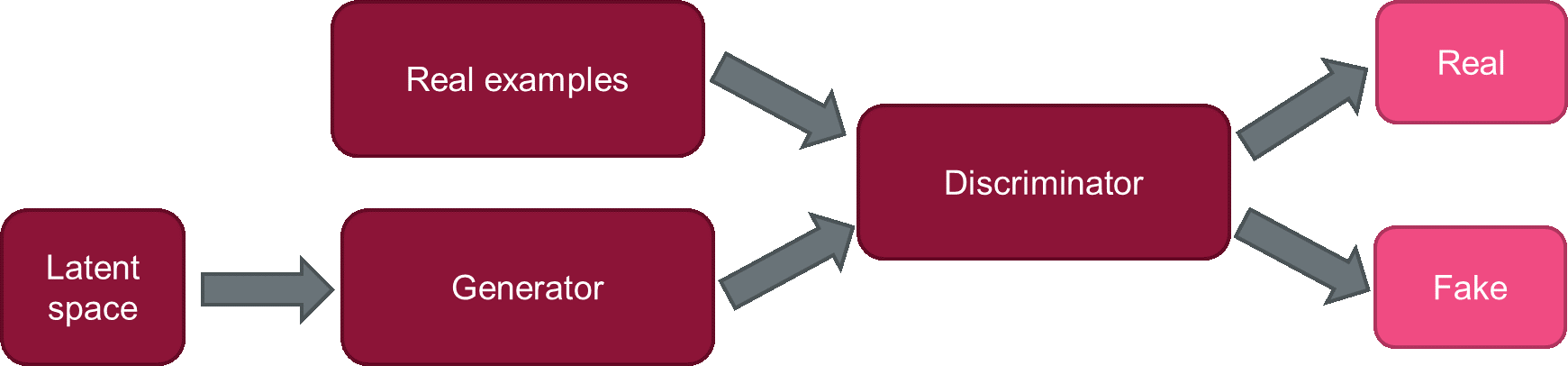}
    \caption{A GAN. The Generator is optimised to produce fakes that fool the
Discriminator. The Discriminator it optimised to identify real examples
from fakes}
    \label{fig:21}
\end{figure}

    A GAN is made up of two deep learning models that compete against each
other. A number of real examples, such as images, are available. As shown in In Figure \ref{fig:21}, the
Generator Model generates new examples and the Discriminator model
classifies the examples as real or fake. Both models are trained. The
Generator is trained to produce fakes that fool the Discriminator. The
Discriminator is trained to identify real examples from fakes.
Ultimately the Generator can produce examples that seem very realistic.
%as seen in Figure \ref{fig:20}.

    In Figure \ref{fig:21} above the latent space is a random seed that ensures that
each Generator example is different.

\begin{mdframed}[style=RedBox, frametitle={%
    \hypertarget{a-quantum-latent-space}{%
\subsubsection{A quantum latent space}\label{a-quantum-latent-space}}}]

    As stated in the introduction ``a possible early use case for quantum
machine learning is to use quantum computers to generate the latent
spaces for GAN models''. A Boson Sampling quantum computer can be used
to generate the latent space, with promising results.

An alternative approach, set out in a 2019
\href{https://arxiv.org/abs/1905.13205}{paper} \cite {Anschuetz2019} adds a Quantum Boltzman machine to the
Discriminator to improve performance.
\end{mdframed}

    \hypertarget{to-conclude}{%
\section{To conclude}\label{to-conclude}}

    This review has described how the ``human like intelligence'' of \textbf{Artificial Intelligence} 
(AI) has evolved from \hyperlink{expertsystems}{\textbf{Expert Systems}} to 
\textbf{Machine Learning} (ML) algorithms that can learn
from data. In a simple \hyperlink{an-introductory-example---linear-regression}{\textbf {linear regression example}} we looked at a \hyperlink{the-hypothesis-space}{\textbf{hypothesis space}} 
of linear predictor functions selected from a large range of possible prediction functions by the \hyperlink{the-inductive-bias}{\textbf{inductive bias}}.

    We defined a \hyperlink{choosing-and-optimising-a-loss-function}{\textbf{loss
function}} to measure the quality of the solution, and observed that many
ML problems are optimisation problems in which parameters are varied to minimise the loss function. We showed
how having too many parameters can lead to
\hyperlink{overfitting}{\textbf{overfitting}}, where the predictions tracked noise in the training
data too closely, and did not generalise properly.

    We then moved from the linear regression \textbf{supervised learning}
example, where labels are provided for each example, to an \textbf{unsupervised learning} \hyperlink{clustering}{\textbf{clustering example}}, where the algorithm
learnt to group the points into clusters without predefined labels for the data.

    Next,  we looked at a \hyperlink{SVC}{\textbf{support vector classifier}} 
which found a linear threshold between two groups of points. 
This example was generalised to a 
\hyperlink{The-support-vector-machine}{\textbf{support vector machine}} 
where the data set, when plotted in two dimensions, 
could not be separated by a linear threshold. Instead the Kernel trick was used to
 introduce a higher dimension in which the data points could be separated by a
hyper-plane.

    Instead of increasing the number of dimensions, in \hyperlink{PCA}{\textbf{Principal Component
Analysis}} (PCA) the number of dimension was reduced. Linear algebra
techniques were used to diagonalise the covariance matrix, and we recognised that the
eigenvectors with the largest eigenvalues (the diagonal entries) 
contained most of the information about the data, allowing a simplified analysis in two or three dimensions.

    Finally, we look at \hyperlink{deep-learning}{\textbf{deep-learning}}, which is inspired by
biology-like networks, and generalised this to a \hyperlink{GAN}{\textbf{Generative Adversarial
Networks}}, where a \textbf{Generator} is locked in a games-theory
inspired competition with a \textbf{Discriminator}. Even a human can struggle to identify the Generator's fakes
from the real images used in training.

    We have also shown where quantum computing might be relevant, and raised
some important \hyperlink{caveats}{\textbf{caveats}} about the readiness of the technology.

    We hope that you have found this review valuable. If you have any
questions or feedback please contact the quantum team at Digital Catapult 
\href{mailto:quantum@digicatapult.org.uk}{\nolinkurl{quantum@digicatapult.org.uk}} 

    \hypertarget{Acknowledgements}{%
\subsection*{Acknowledgements}\label{Acknowledgements}}

    Many thanks to Robert Elliott Smith for his helpful comments.

    \hypertarget{Appendix1}{%
\section*{Appendix - The mathematics of Principal Component Analysis}\label{Appendix - The mathematics of Principal Component Analysis}}

This Appendix explains how the covariance matrix is diagonalised, and the principal components are identified as the eigenvectors with the largest eigenvalues.

Following
\href{https://mml-book.github.io/book/mml-book.pdf}{Mathematics for
Machine Learning} \cite {MML} we consider a set of \(N\) data vectors \(x_n\), each
of dimension \(D\), with mean zero. The dataset can be written
\(\mathscr{X} = \left\{{\pmb x_1,,\pmb x_n,,\pmb x_N}\right\}\) where
\(\pmb x_n \in\ \mathscr{R}^D\). Because the mean is zero the data
covariance matrix is a $D \times D$ matrix defined as:

    \[ \pmb S = \frac{1}{N} \sum_{n=1}^{N} \pmb x_n \pmb x_n^T\]

    The data covariance matrix is a symmetric matrix. The diagonal elements
represent the variance, a measure of how much variability there is in
each dimension of the data. The other matrix elements represent the
covariances, a measure of how the variability in one data dimension is
linked to variablity in the other. This matrix is very similar to the
density matrix $\rho$ in quantum mechanics where the sum of the diagonals is
normalised to one, and the off-diagonal terms hold details of
correlations: \[\rho =  |\psi \rangle \langle \psi|\]

    There is a well known theorem from linear algebra that a symmetric
matrix like $\pmb S$ can be diagonalised and we can always write:

    \[\pmb S = \pmb P \pmb D \pmb P^{-1} \]

    \(\pmb D\) is a diagonal matrix and it is often the case that some of
the diagonal entries are much larger than others. Conventionally
\(\pmb D\) is written with the largest eigenvector \(\lambda_1\) at the
top, and the rest in descending order:

    \begin{equation*}
\pmb D = 
\begin{bmatrix}
& \lambda_1 & 0 & . & . & .\\
& 0 & \lambda_2 & 0 & . & .\\
& 0 & 0 & \lambda_3 & . & .\\
& . & . & . & . & . \\
& 0 & . & . & . & \lambda_D
\end{bmatrix}
\end{equation*}

    If we multiply both sides of the equation
\(\pmb S = \pmb P \pmb D \pmb P^{-1}\) above by \(\pmb P\) from the
right then we get.

    \[ \pmb S \pmb P = \pmb D \pmb P\]
 and considering the column vectors of \(\pmb P\) : 
\(\pmb p_1, \pmb p_2, \pmb p_M\) then we see these satisfy an eigenvector equation:

    \[\pmb S \pmb p_i = \lambda_i \pmb p_i\]

    The column vectors \(\pmb p_i\) are all eigenvectors of \(\pmb S\).
Those associated with the largest diagonal entries are the
\textbf{principal components}. Moreover these eigenvectors are row
vectors of \(\pmb P^{-1}\) since it is known that is known that
\(\pmb P^{T} = \pmb P^{-1}\).

\begin{figure}[h]
    \centering
    \includegraphics[width=1\textwidth]{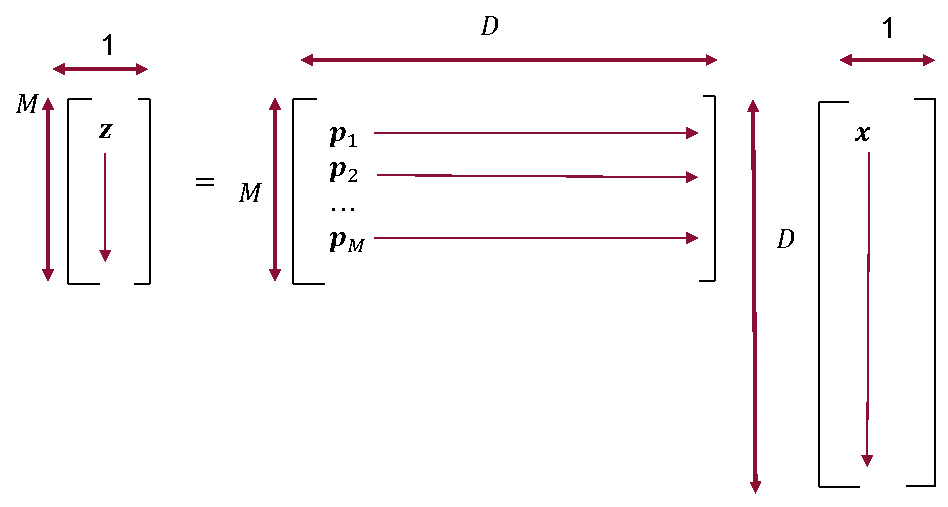}
    \caption{A vector \(\pmb z\) in reduced M dimensional space can be produced
by multiplying a D dimensional data vector \(\pmb x\) by the transpose
of the first M eigenvectors of the covariance matrix.}
    \label{fig:22}
\end{figure}

    If we take the first M rows of \(\pmb P^{-1}\) this \(M X N\) matrix
will project the data onto a lower dimension \(M x M\) subspace whilst
retaining much of the information. The matrix multiplication is shown
in Figure \ref{fig:22}

\begin{small}
\bibliographystyle{unsrt} % We choose the "unsorted" reference style like Hassan
 %\bibliography{Review_of_ML_arxiv.bib}
\bibliography{Review_of_ML_arxiv}
\end{small}

\end{document}